\documentclass{aa}

\usepackage[varg]{txfonts}
\usepackage{siunitx}
\usepackage[version=4]{mhchem}
\usepackage{multirow}
\usepackage{graphicx}
\usepackage[caption=false]{subfig}
\usepackage[strict]{changepage}
\usepackage{xcolor}
\usepackage{ulem}
\usepackage[colorlinks=true,citecolor=blue]{hyperref}
\usepackage{float}
\usepackage{amssymb}
\usepackage{amsmath}
\usepackage{lscape}
\usepackage{mathtools}
\usepackage{natbib}
\usepackage{pdflscape}
\usepackage{gensymb}

\defcitealias{zhou2023}{Paper~I}   

\DeclareSIUnit\jansky{Jy}

\newcolumntype{L}[1]{>{\raggedright\let\newline\\\arraybackslash\hspace{0pt}}m{#1}}
\newcolumntype{C}[1]{>{\centering\let\newline\\\arraybackslash\hspace{0pt}}m{#1}}
\newcolumntype{R}[1]{>{\raggedleft\let\newline\\\arraybackslash\hspace{0pt}}m{#1}}

\begin{document}

\title{High-resolution APEX/LAsMA $^{12}$CO and $^{13}$CO (3-2) observation of the G333 giant molecular cloud complex : III. Decomposition of molecular clouds into multi-scale hub-filament structures}

\author{J. W. Zhou\inst{\ref{inst1}} 
\and J. S. Urquhart\inst{\ref{inst2}}
\and F. Wyrowski \inst{\ref{inst1}}
\and S. Neupane \inst{\ref{inst1}}
\and T. Liu \inst{\ref{inst3}}
\and M. Juvela \inst{\ref{inst4}}
}
\institute{
Max-Planck-Institut f\"{u}r Radioastronomie, Auf dem H\"{u}gel 69, 53121 Bonn, Germany \label{inst1} \\
\email{jwzhou@mpifr-bonn.mpg.de}
\and
Centre for Astrophysics and Planetary Science, University of Kent, Canterbury, CT2 7NH, UK \label{inst2}\\
\and
Shanghai Astronomical Observatory, Chinese Academy of Sciences, 80 Nandan Road, Shanghai 200030, Peoples Republic of China \label{inst3}\\
\and
Department of Physics, P.O.Box 64, FI-00014, University of Helsinki, Finland \label{inst4}\\
}

\date{Accepted XXX. Received YYY; in original form ZZZ}

\abstract
{We decomposed the G333 complex and the G331 giant molecular cloud into multi-scale hub-filament systems (HFs) using the high-resolution $^{13}$CO (3$-$2) data from LAsMA observations. We employed the filfinder algorithm to identify and characterize filaments within HFs. Compared with non-HFs, HFs have significantly higher density contrast, larger masses and lower virial ratios. Velocity gradient measurements around intensity peaks provide evidence of gas inflow within these structures.
There may be an evolutionary sequence from non-HFs to HFs. Currently, non-HFs lack a distinct gravitational focusing process that would result in significant density contrast.  
The density contrast can effectively measure the extent of gravitational collapse and the strength of the gravitational center of the structure that definitively shape the hub-filament morphology. 
Combined with the kinematic evidence in our previous studies, we suggest that molecular clouds are network structures formed by the gravitational coupling of multi-scale hub-filament structures. The knots in the networks are the hubs, they are the local gravitational centers and the main star-forming sites. 
Actually, clumps in molecular clouds are equivalent to the hubs. 
The network structure of molecular clouds can naturally explain that feedback from protoclusters does not significantly change the
kinematic properties of the surrounding embedded dense gas structures, as concluded in our previous studies.}


\titlerunning{Decomposition of molecular clouds into multi-scale hub-filament structures}
\authorrunning{J. W. Zhou}

\maketitle 

\section{Introduction}\label{intro}
Assessing the dynamic interaction between density fluctuations in molecular clouds and the surrounding gas motions offers insight into the development of hierarchical structures within high-mass star-forming regions
\citep{McKee2007-45, Schisano2014-791, Motte2018-56, Henshaw2020-4,Hacar2023-534}. 
High-resolution observations of high-mass star-forming regions show that density enhancements are organized in filamentary gas networks, particularly in hub-filament systems. 
Hub-filament systems  are known as a junction of three or more filaments. Filaments have lower column densities compared to the hubs \citep{Myers2009,Schneider2012,Kumar2020-642,Zhou2022-514}. 
In these systems, converging flows channel matter into the hub via filaments. Earlier research has proposed that hub-filament structures serve as the nurseries for the formation of high-mass stars and clusters
\citep{Peretto2013,Henshaw2014,Zhang2015,Liu2016,Yuan2018,Lu2018,Issac2019,Dewangan2020,Liu2021-646,Liu2022-511,Kumar2020-642,Zhou2022-514,Liu2023-522,Xu2023-520,Zhou2023-676,Yang2023-953,Zhou2024-686-146}. 
Statistically, \citet{Kumar2020-642} identified the candidate hub-filament systems in the Milky Way inner disc by searching for filamentary structures around 35 000 clumps in the Hi-GAL catalogue \citep{Elia2017-471}. $\sim$11\% of the clumps are the candidates of hub-filament systems. The column densities of hubs are found to be enhanced from a factor of approximately two (pre-stellar sources) up to about ten (proto-stellar sources).  High-mass stars preferentially form in the density-enhanced hubs of hub-filament systems. This amplification can drive the observed longitudinal flows along filaments providing further mass accretion.

Specifically, \citet{Zhou2022-514} investigated the physical characteristics and evolution of hub-filament systems in approximately 140 protoclusters using spectral line data from the ATOMS (ALMA Three-millimeter Observations of Massive Star-forming regions) survey \citep{Liu2020}. They suggested that hub-filament structures, characterized by self-similarity and filamentary accretion, remain consistent across various scales in high-mass star-forming regions, ranging from several thousand astronomical units to several parsecs. This model of hierarchical, multi-scale hub-filament structures was extended from the clump-core scale to the cloud-clump scale by \citet{Zhou2023-676}, and further developed to the galaxy-cloud scale in \citet{Zhou2024arXiv240415862Z,Zhou2024-534}. Earlier studies have also illustrated hierarchical collapse and hub-filament structures feeding central regions, as described by \citet{Motte2018-56}, \citet{Vazquez2019-490} and \citet{Kumar2020-642}, and references therein.

Kinematically, the results in \citet{Zhou2023-676} have revealed the multi-scale hub-filament structures in the G333 complex. 
Numerous structures within the G333 complex exhibit the characteristic kinematic properties of hub-filament systems. For instance, the intensity peaks, which represent the hubs, are linked to converging velocities, indicating that the surrounding gas is flowing towards and accumulating in these dense regions.
In particular, the velocity gradients increase at small scales. The filaments in both LAsMA (Large APEX sub-Millimeter Array) and ALMA observations show clear velocity gradients. The velocity gradients we fit to the LAsMA and ALMA data agree with each other over the range of scales covered by ALMA observations in the ATOMS survey ($\textless$ 5\,pc). In \citet{Zhou2023-676}, larger-scale gas motions were investigated, the longest filament being $\sim$50\,pc, but we obtained similar results to small-scale ALMA observations. The variations in velocity gradients at small scales ($\textless$ 1\,pc), medium scales ($\sim$ 1-7.5\,pc), and large scales ($\textgreater$ 7.5\,pc) are consistent with gravitational free-fall with central masses of $\sim$ 500\,M$_\odot$, $\sim$ 5000\,M$_\odot$ and $\sim$\, 50000\,M$_\odot$. This means that the velocity gradients on larger scales require a higher mass to be maintained. 
Higher masses correspond to larger scales, indicating that large-scale inflow is driven by larger-scale structures. This may result from the gravitational clustering of smaller-scale structures, aligning with the hierarchical nature of molecular clouds and the gas inflow from larger to smaller scales. The large-scale gas inflow is likely driven by gravity, suggesting that the molecular clouds in G333 are undergoing global gravitational collapse. 
Additionally, the funnel-like velocity field observed in position-position-velocity (PPV) space further supports the concept of large-scale gravitational collapse within the molecular clouds of the G333 complex \citep{Zhou2023-676}.
The change in velocity gradients with the scale in the G333 complex indicates that the morphology of the velocity field in PPV space resembles a "funnel" structure \citep{Zhou2023-676}. The funnel structure can be explained as accelerated material flowing towards the central hub and gravitational contraction of star-forming clouds/clumps. Large-scale velocity gradients always involve many intensity peaks, and the larger-scale inflow is driven by the larger-scale structure, implying that the clustering of local small-scale gravitational structures can act as the gravitational center on a larger scale. To some extent, the funnel structure gives an indication of the gravitational potential well that is formed by the clustering \citep{Zhou2023-676}. The “funnel” behaviour in terms of gravitational collapse has also been described in previous works, such as \citet{Kuznetsova2015-815} and \citet{Hacar2016-589,Hacar2017-602}. Based on the evidence from the kinematics, the main goal of this work is to directly recover the multi-scale hub-filament structures in the G333 complex and the G331 giant molecular cloud (GMC) using the high-resolution $^{13}$CO (3$-$2) data in LAsMA observation. The structure of this paper is as follows. We describe the data sources in Sec.\ref{data}. We present our main results in Sec.\ref{result}, and discuss their implications or inferences in Sec.\ref{discuss}. We summarize our findings in Sec.\ref{summer}.

\section{Data}\label{data}
The observations and
data reduction are described in detail in \citet{Zhou2023-676}.
We mapped a $3.4\degr \times 1.2\degr$ area centered at $(l,b)=(332.33\degr,-0.29\degr)$ using the APEX telescope \citep{Gusten2006-454}, which includes the G333 complex and the G331 GMC.
\footnote{\tiny This publication is based on data acquired with the Atacama Pathfinder EXperiment (APEX) while it has been a collaboration between the Max-Planck-Institut f\"{u}r Radioastronomie, the European Southern Observatory and the Onsala Space Observatory.}
The 7 pixel LAsMA receiver was used to observe the $J=3-2$ transitions of $^{12}$CO ($\nu_{\text{rest}}\sim345.796$ \rm GHz) and $^{13}$CO ($\nu_{\text{rest}}\sim 330.588$ \rm GHz) simultaneously. 
Observations were performed in a position switching on-the-fly (OTF) mode. 
The data were reduced using the GILDAS package\footnote{\url{http://www.iram.fr/IRAMFR/GILDAS}}. The final data cubes have a velocity resolution of 0.25 km s$^{-1}$, an angular resolution of $19.5 \arcsec$ and a pixel size of $6 \arcsec$.
A main-beam efficiency value $\eta_{\rm mb}=0.71$ \citep{Mazumdar2021-650} was used to convert intensities from the $T^*_A$ scale to main beam brightness temperatures, $T_{\rm mb}$. The resolution of the CO data is consistent with that of the ATLASGAL survey \citep{Schuller2009-504}. In this work, we focus on hub-filament structures on cloud-clump scales (1–10 pc). As shown in Sec.\ref{assoc}, most of the hubs are associated with clumps identified in the ATLASGAL survey.

The observed region was covered in the IR range by the Galactic Legacy Infrared Midplane Survey \citep[GLIMPSE,][]{Benjamin2003}.
GLIMPSE images, obtained with the Spitzer Infrared Array Camera (IRAC) at 8.0 $\mu$m, were retrieved from the Spitzer archive. The angular resolution of the images in the IRAC bands is $\sim 2\arcsec$.

\section{Results}\label{result}

\subsection{Identification of intensity peaks}\label{dendrogram}

\begin{figure*}[htbp!]
\centering
\includegraphics[width=0.95\textwidth]{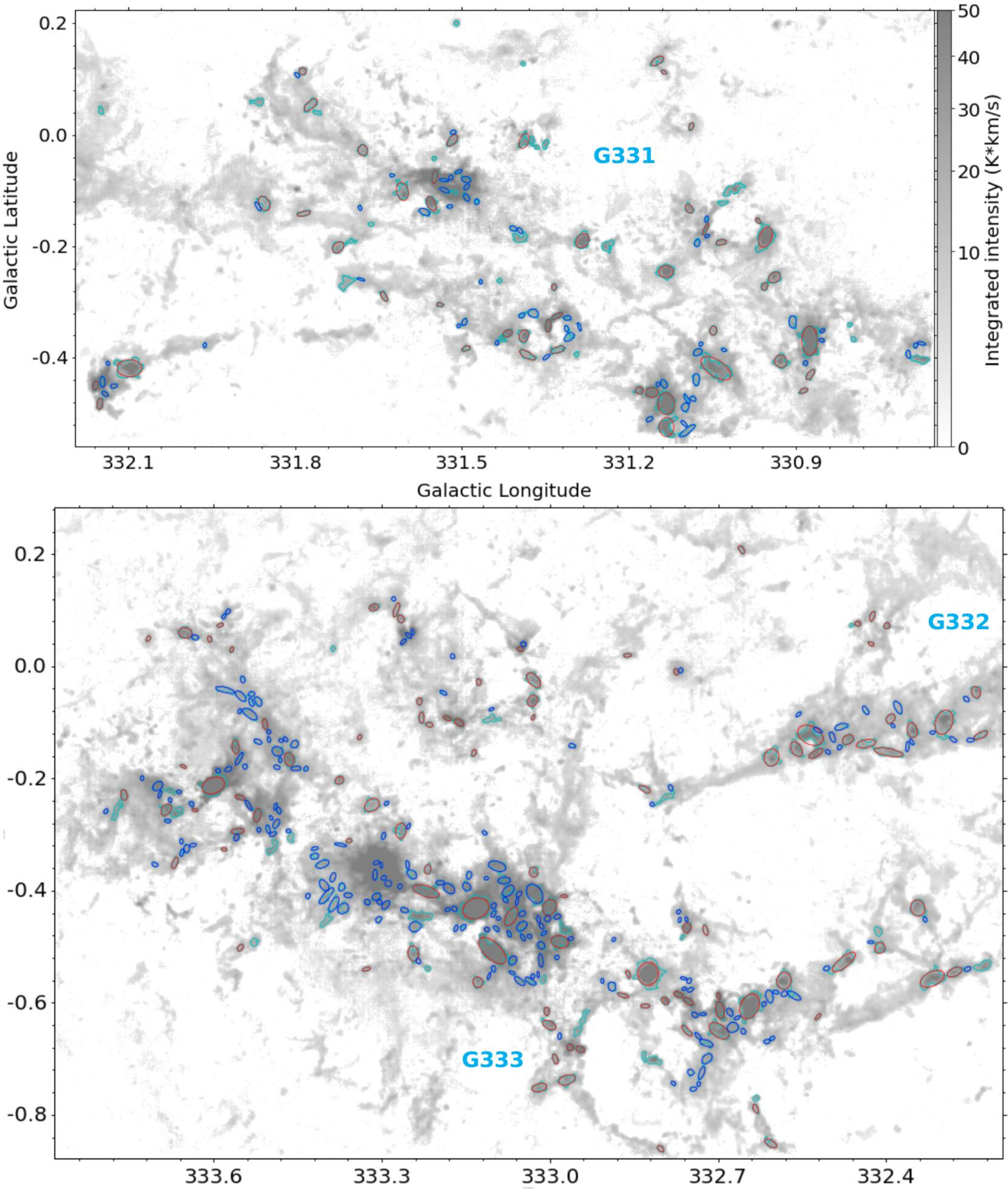}
\caption{The background is the integrated intensity map of $^{13}$CO (3$-$2) in the full velocity range [$-$120, $-$20] km s$^{-1}$, that is covered by the masks of leaf structures (cyan contours) identified by the dendrogram algorithm.
The central regions of non-HFs and the hubs of HFs, as classified in Sec.\ref{class}, are marked by blue and red ellipses, respectively.}
\label{large}
\end{figure*}

The most likely hub candidates are the intensity peaks. Following the procedure described in \citet{Zhou2024-682-128}, we directly identify the intensity peaks based on the integrated intensity (Moment 0) map of $^{13}$CO (3$-$2) emission. 
Using the {\it astrodendro} package \footnote{\url{https://dendrograms.readthedocs.io/en/stable/index.html}},
there are three major input parameters for the dendrogram algorithm: {\it min\_value} for the minimum value to be considered in the dataset, {\it min\_delta} for a leaf that can be considered as an independent entity, and {\it min\_npix} for the minimum area of a structure. The dendrogram algorithm decomposes the intensity data into hierarchical structures, called leaves and branches.
For the Moment 0 map, a 5$\sigma$ threshold has been set. In \citet{Zhou2024-682-128}, we only require the minimum structural area to exceed the size of a single beam and avoid setting additional parameters, thus minimizing the algorithm's reliance on specific parameter settings for structure identification.
In this work, we focus on the hubs and are mainly interested in the strongest local intensity peaks. In addition to setting {\it min\_npix} as one beam, a threshold 15 K km s$^{-1}$ was also adopted for the parameter {\it min\_value} to filter the relatively low-density structures. 
We tried different values of {\it min\_value}, and found the value of 15 K km s$^{-1}$ can best recover all high intensity peaks, as shown in Fig.\ref{large}. 
Finally, a total of 612 structures are identified, including 438 leaves and 174 branches. In this work, we only focus on leaf structures (leaves). Then, we applied the fully automated Gaussian decomposer algorithm, \texttt{GAUSSPY+} \citep{Lindner2015-149, Riener2019-628}, to fit the averaged spectra of leaf structures individually. The parameter settings for the decomposition were identical to those used by \citet{Zhou2023-676}.

{\it astrodendro} approximates the morphology of each structure as an ellipse. In the dendrogram, the long and short axes of an ellipse, $a$ and $b$, represent the root mean square (rms) sizes (second moments) of the intensity distribution along the two spatial dimensions. As outlined in \citet{Zhou2024-682-128}, these rms sizes result in a smaller ellipse compared to the actual size of the identified structure. Therefore, it is necessary to multiply these dimensions by a factor of 2 to properly scale the ellipse to match the structure's size.
The effective physical radius of an ellipse is then $ R\rm_{eff}$ =$\sqrt{2a \times 2b} \times d$, with $d=3.6~{\rm kpc}$ for the distance to the G333 complex \citep{Lockman1979-232,Bains2006-367}.

In Fig.2 of \citet{Zhou2023-676}, the average spectrum of $^{13}$CO (3$-$2) for the entire region (the G333 complex and the G331 GMC) shows that the emission is concentrated in the velocity range [$-$120, $-$20] km s$^{-1}$, and there are three main velocity components peak1, peak2 and peak3 defined in \citet{Zhou2023-676}. In \citet{Zhou2023-676}, we only considered the peak3, which is regarded as the G333 molecular cloud complex. However, in Fig.4(b) of \citet{Zhou2023-676}, a large-scale velocity gradient links peak 3 with peak 2, forming a loop structure between them.
In \citet{Nguyen2015-812}, both peak2 and peak3 were assigned to the same molecular cloud complex in the Scutum-Centaurus arm with the distance $\sim 3.6~{\rm kpc}$, while peak1 is thought to be located in the Norma arm at a distance $\sim 5~{\rm kpc}$. However, the associations between peak1, peak2 and peak3 are still unclear. Four ATLASGAL clump clusters marked in Fig.3 of \citet{Zhou2023-676} were distributed to similar distances $\sim 3.6~{\rm kpc}$ in \citet{Urquhart2022-510}, although one of them belongs to peak1. In Fig.2 and Fig.3 of \citet{Zhou2023-676}, peak1 and peak2 were called G331 GMC-blue and -red, because they seem to be two symmetrical halves of the G331 GMC. However, as shown in Fig.2 of \citet{Zhou2023-676}, their velocity difference is $\sim$ 25 km s$^{-1}$, thus they are put in different spiral arms in \citet{Nguyen2015-812}. In this work, we consider the entire velocity range [$-$120, $-$20] km s$^{-1}$, i.e. both the G333 complex and the G331 GMC, and take the  distance $d=3.6~{\rm kpc}$ of the G333 complex \citep{Lockman1979-232,Bains2006-367} for all identified structures. Considering the emission of peak2 and peak3 is dominant in the entire observed region, therefore, the potential distance bias of peak1 will not significantly affect the statistical results. Moreover, there is no big difference whether the distance is $\sim 5~{\rm kpc}$ or $\sim 3.6~{\rm kpc}$ for peak1.

\subsection{Velocity components}\label{component}

\begin{figure*}
\centering
\includegraphics[width=1\textwidth]{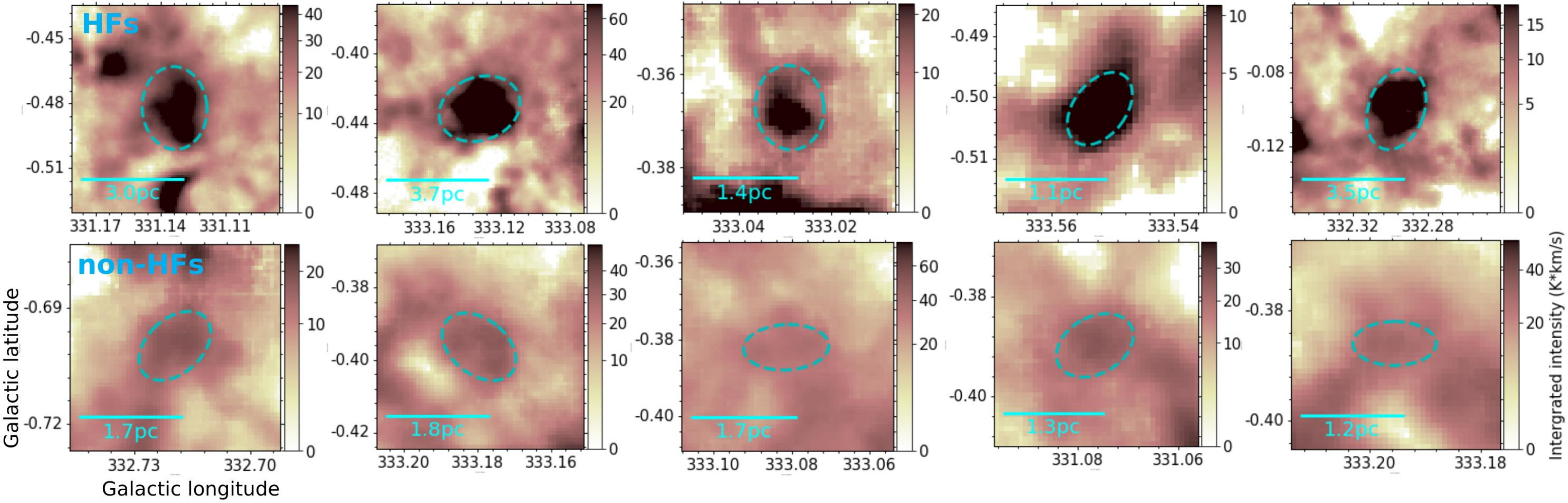}
\caption{Some examples of HFs and non-HFs classified in Sec.\ref{class}. 
The dashed ellipses represent the equivalent ellipses of the identified leaf structures.
The size of the boxes is 2.5 times that of the hub (defined as the effective diameter of the corresponding leaf structure marked by the ellipse), reflecting the spatial range used to extract the average spectra presented in Fig.\ref{typical}.}
\label{case}
\end{figure*}

\begin{figure}
\centering
\includegraphics[width=0.5\textwidth]{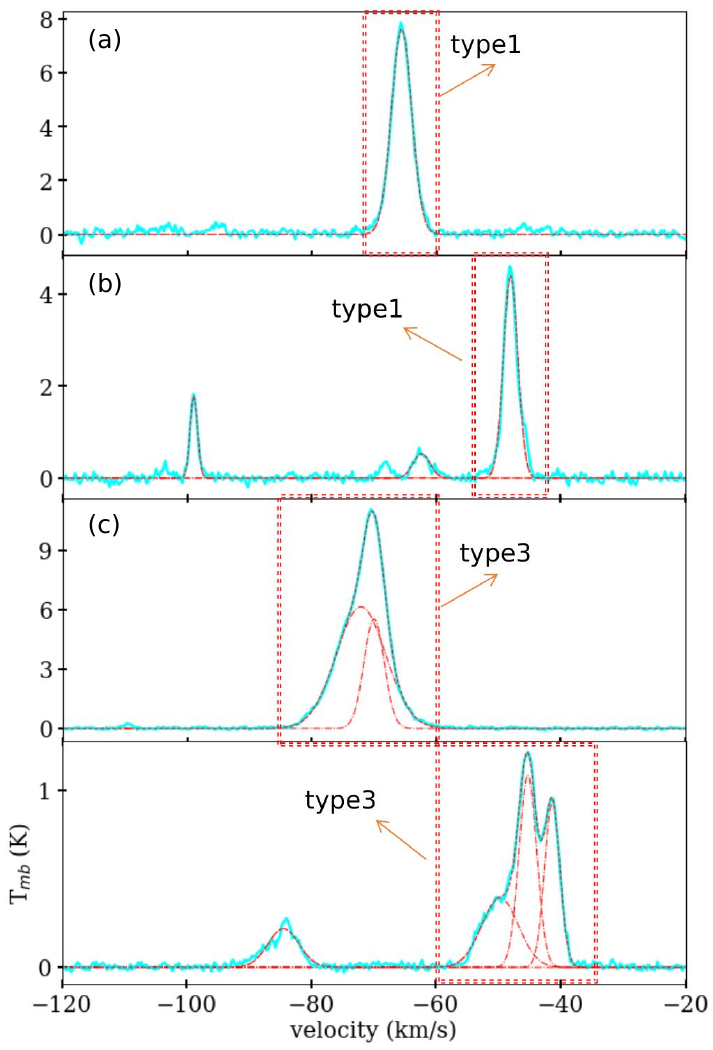}
\caption{Typical $^{13}$CO (3$-$2) line profiles of the identified structures. Red dashed box marks the limited velocity range of each structure.}
\label{typical}
\end{figure}

For the hub-filament systems, as presented in the literature listed in Sec.\ref{intro}, hubs are essentially structures with higher density compared to the surrounding diffuse gas. As local gravitational centers, these hubs can accrete the surrounding diffuse gas, forming hub-filament structures. The filamentary structures are actually gas inflows converging towards the hubs. A hub-filament structure essentially consists of a gravitational center and gas inflows converging towards it. Therefore, all identified intensity peaks are thought to be potential hubs. Around the intensity peaks, we extended the spatial ranges to investigate the filamentary structures connected with them. As shown in the first row of Fig.\ref{case}, extending to 2.5 times the hub size (the effective radius of the corresponding leaf structure) can well recover a relatively complete hub-filament morphology. However, the hub-filament structures are identified directly according to the integrated intensity map of the full velocity range. Therefore, the issue of the overlapping of different velocity components is inevitable, as shown in Fig.3 of \citet{Zhou2023-676}. To address this issue, we extracted the average spectra for all structures in the 2.5 times extended regions to decompose their velocity components.

The velocity components within the same structure that exhibit large velocity differences are generally independent of each other, which can be attributed to projection effects.
Thus, for each structure, we mainly focus on the main emission marked by the red dashed box in Fig.\ref{typical}. As shown in Fig.\ref{typical}(a) and (c), the main emission can be one velocity component or multiple blended velocity components. Using the names in \citet{Zhou2024-682-128}, they are called type1 and type3 structures. For type2 structures with separated velocity components (such as Fig.\ref{typical}(b)), we only consider the dominant component, thus they are reduced to type1 structures. 

The velocity range of a structure is essential for accurately estimating its fundamental physical properties, including velocity dispersion and mass.
As discussed in \citet{Zhou2024-682-128},
we take the velocity range of each velocity component as [v$_{c}$-FWHM,v$_{c}$+FWHM]. For a type3 structure, the velocity range for the leftmost and rightmost velocity components are [v$_{c1}$-FWHM$_{1}$,v$_{c1}$+FWHM$_{1}$] and [v$_{c2}$-FWHM$_{2}$,v$_{c2}$+FWHM$_{2}$], respectively. Then the total velocity range of the type3 structure is 
[v$_{c1}$-FWHM$_{1}$,v$_{c2}$+FWHM$_{2}$].

In the first row of Fig.\ref{case}, the hub-filament structures are connected to the gas environment by the filamentary structures. They are not independent, as the filamentary structures lack a clear boundary with the surrounding gas. Therefore, accurately estimating the physical properties of the entire hub-filament structures, such as their scale and density, is challenging.
But the hubs are relatively isolated, and as local structures, it is feasible to calculate their physical parameters. 
However, the velocity range of the hub may be different from that of the entire hub-filament structure.
Thus, we repeat the procedures described above to estimate the velocity ranges and velocity dispersion of the hubs. 
Following the method outlined in \citet{Zhou2024-682-128}, we extract the physical quantities for each hub by utilizing its velocity range and effective ellipse based on the column density ($N$) and temperature cubes
derived from $^{13}$CO (3-2) and $^{12}$CO (3-2) lines by the local thermodynamic equilibrium (LTE) analysis in the full velocity range [$-$120, $-$20] km s$^{-1}$ \citep{Zhou2023-676}. Then we calculate the mass and virial ratio of each structure, as done in \citet{Zhou2024-682-128}.

In this work,
we consider type3 structures to be independent entities, similar to type1 structures, but characterized by more complex gas motions. 
If the different velocity components within a type3 structure are parts of the same structure, the velocity dispersion of the type3 structure can be determined by calculating an intensity-weighted average of the velocity dispersion of all its components. 

Evidence supporting the hypothesis that a type3 structure is a single entity, rather than a result of overlapping uncorrelated gas components, includes the following points:
(1) In Fig.\ref{typical}, the significant overlap in the line profiles of type3 structures may indicate a correlation between different velocity components within the structure. The greater the overlap between these components, the more likely they are part of the same structure. In extreme cases, this overlap results in a single-peak line profile (type1 structure);
(2) This study focuses on hub-filament structures, where gas inflows along the filaments exhibit varying velocities, leading to multiple velocity components within the same structure;
(3) In Sec.\ref{filament}, for type3 structures,
the velocity variation along the filament is continuous, without abrupt changes.

\subsection{Classification}\label{class}

\begin{figure*}
\centering
\includegraphics[width=0.8\textwidth]{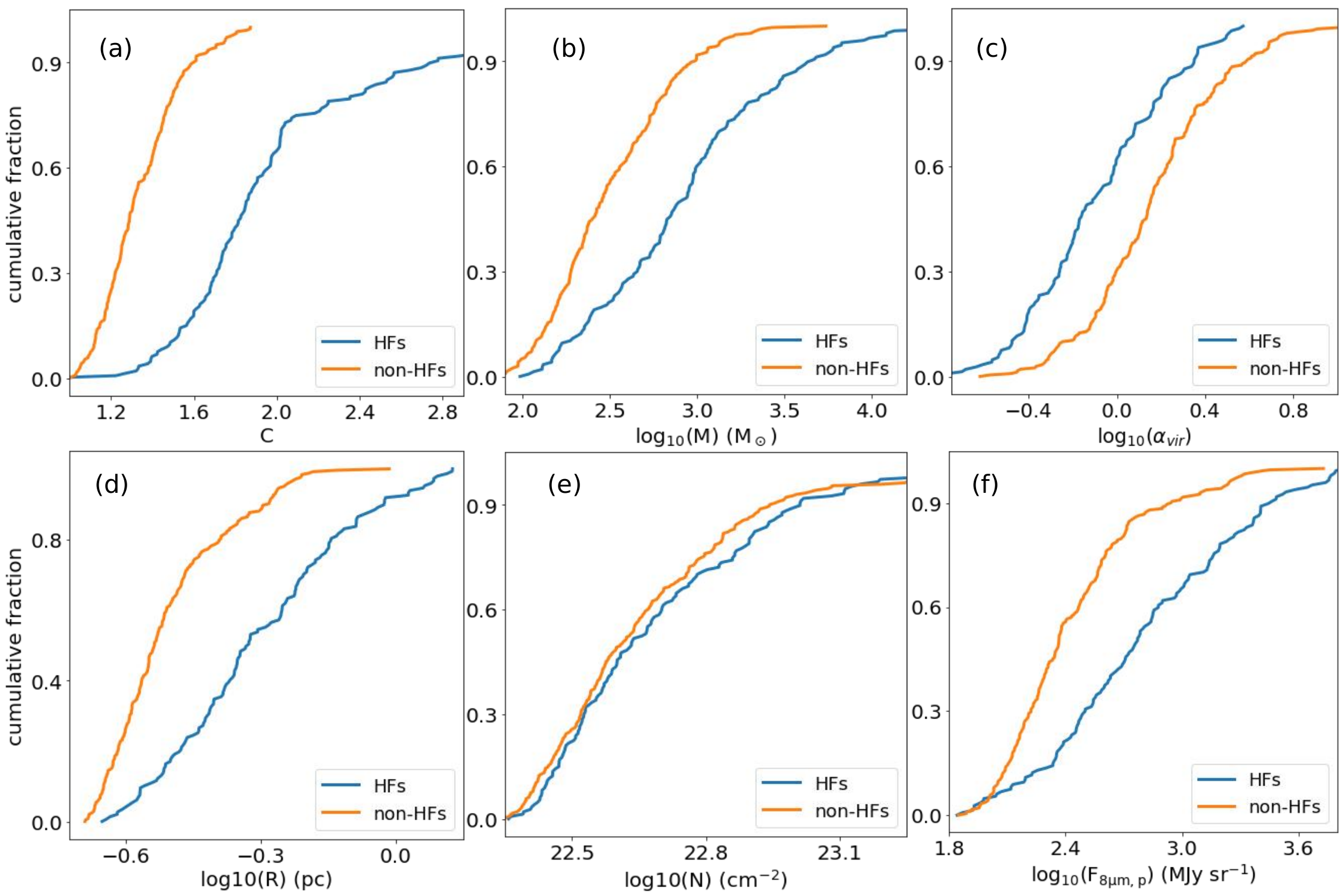}
\caption{Physical properties of HFs and non-HFs. (a) Column density contrast defined in Sec.\ref{contrast}; (b) Mass; (c) Virial ratio; (d) Effective radius; (e) Average column density; (f) Peak intensity of {\it Spitzer} 8 $\mu$m emission.}
\label{compare}
\end{figure*}

Restricted to the fixed velocity ranges of all structures (2.5 times extended) in Sec.\ref{component}, their Moment-0 maps were reproduced. 
In hub-filament systems (HFs), hubs exhibit substantially higher densities compared to filaments, which means a significant density contrast between the hub and the surrounding diffuse gas. We define the density contrast $C$ as the ratio of the average column density in the hub region, $N_{\rm hub}$, to the average column density within an elliptical ring surrounding the hub, which has a width equal to the radius of the hub (the effective radius of the leaf structure), $N_{\rm a}$,
\begin{equation}
C=N_{\rm hub}/N_{\rm a}. 
\end{equation}
According to the morphology and the $C$ value, we categorized the structures into two groups: those with a distinct high-density central region and those without. 
Generally, $C\approx1.5$ can serve as the boundary between the two categories. However, it is difficult to classify structures with $C\approx1.5$ based on morphology. As a preliminary result,
the two categories comprise 195 and 243 structures (non-HFs), respectively.
For the 195 structures featuring a distinct high-density center, we further identified the associated filaments as described in Sec.\ref{filament}. Structures in which filaments converge at the high-density center were classified as hub-filament systems, totaling 148 structures (HFs). 
A small number of remaining structures, which were challenging to classify, were excluded to maintain the clarity of comparison in the subsequent analysis.
Fig.\ref{case} illustrates examples of non-HFs and HFs. In Fig.\ref{compare}, we present a comparison of the physical properties of HFs and non-HFs, revealing significant differences between the two groups.

\subsection{Filamentary structures}\label{filament}

\subsubsection{Identification}

\begin{figure*}
\centering
\includegraphics[width=0.95\textwidth]{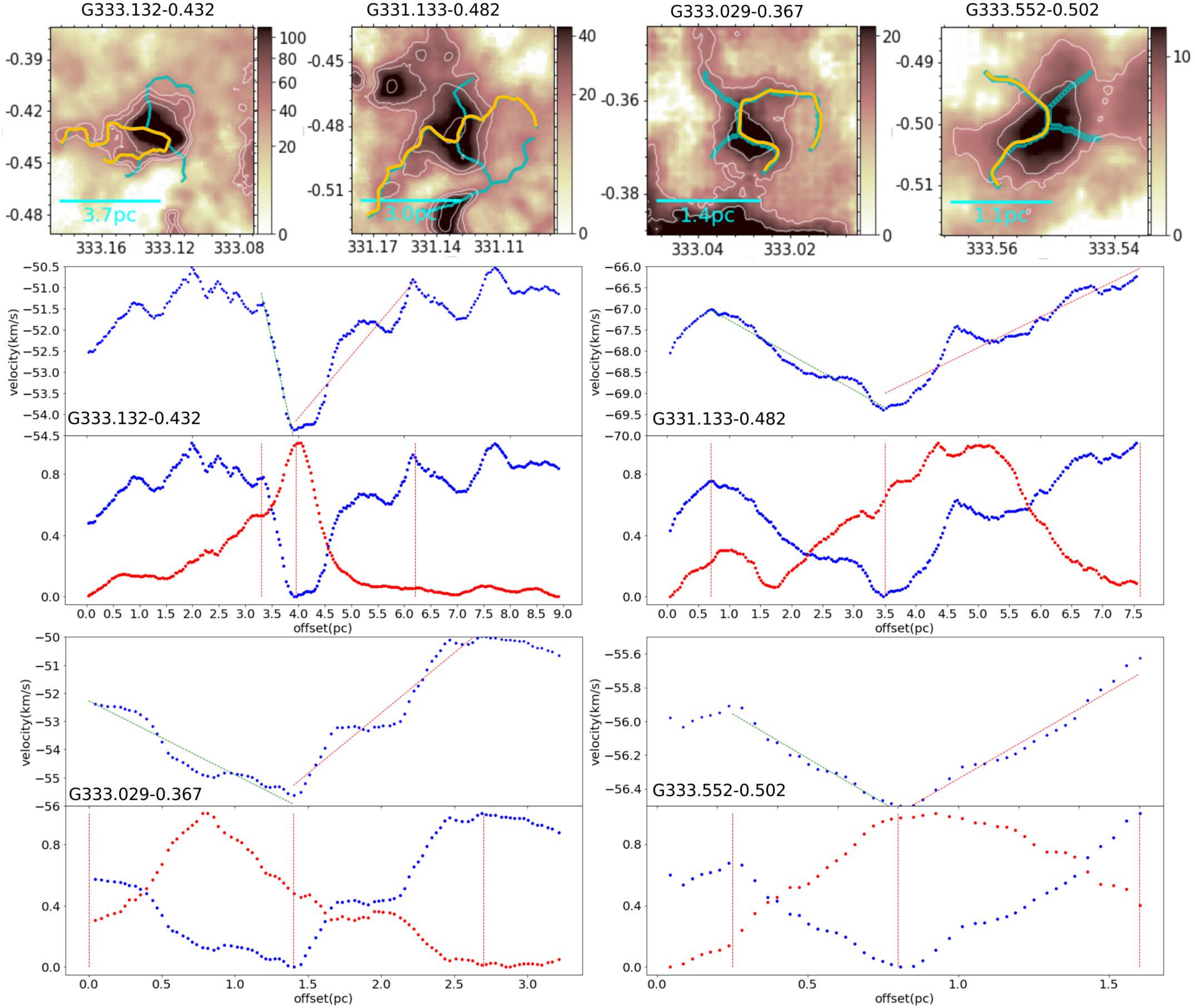}
\caption{Four HFs structures used to demonstrate the velocity gradient fitting. The beamsize is $\sim$0.34 pc.
In the first row, the background is the integrated intensity map of $^{13}$CO (3$-$2), orange and cyan lines are the filaments identified by the filfinder algorithm. 
The background images were regridded, in contrast to those shown in Fig.~\ref{case}.
The two rows below show the velocity and intensity profiles extracted along the orange filaments. In panel (a), velocity gradients are fitted in the ranges defined by the red vertical dashed lines in panel (b). Straight lines show the linear fitting results.
In panel (b), blue and red dotted lines show the normalized velocity and intensity, respectively.}
\label{case-g}
\end{figure*}

As done in \citet{Zhou2023-676}, we also use the filfinder algorithm \citep{Koch2015-452} to characterize the filamentary structures around the hub. 
In \citet{Zhou2023-676}, our focus was on large-scale filaments within molecular clouds. Along these filaments, we observed clear fluctuations in both velocity and intensity, with each intensity peak indicating a local hub. In this study, we began by identifying local intensity peaks (i.e. leaf structures), and subsequently searched for hub-filament structures centered around these intensity peaks. Our current objective is to identify and analyze small-scale filamentary structures within the local hub-filament systems. However, for small-scale structures, the number of pixels is insufficient to directly identify filaments using the filfinder algorithm. To address this, we regridded the images, increasing the pixel count from $N_{x}*N_{y}$ to $2N_{x}*2N_{y}$ \footnote{This was achieved using the \textbf{zoom} function from the \textbf{scipy.ndimage} module, with an interpolation order of 3 for regridding. Regridding only increases the number of pixels; it does not change the image resolution.}. To minimize the introduction of artificial structures, we only doubled the number of pixels. As shown in Fig.~\ref{case} and Fig.~\ref{case-g}, the morphology of the structures remains unchanged. The identified filamentary structures also align well with the background.

\subsubsection{Velocity gradient}

\begin{figure}
\centering
\includegraphics[width=0.45\textwidth]{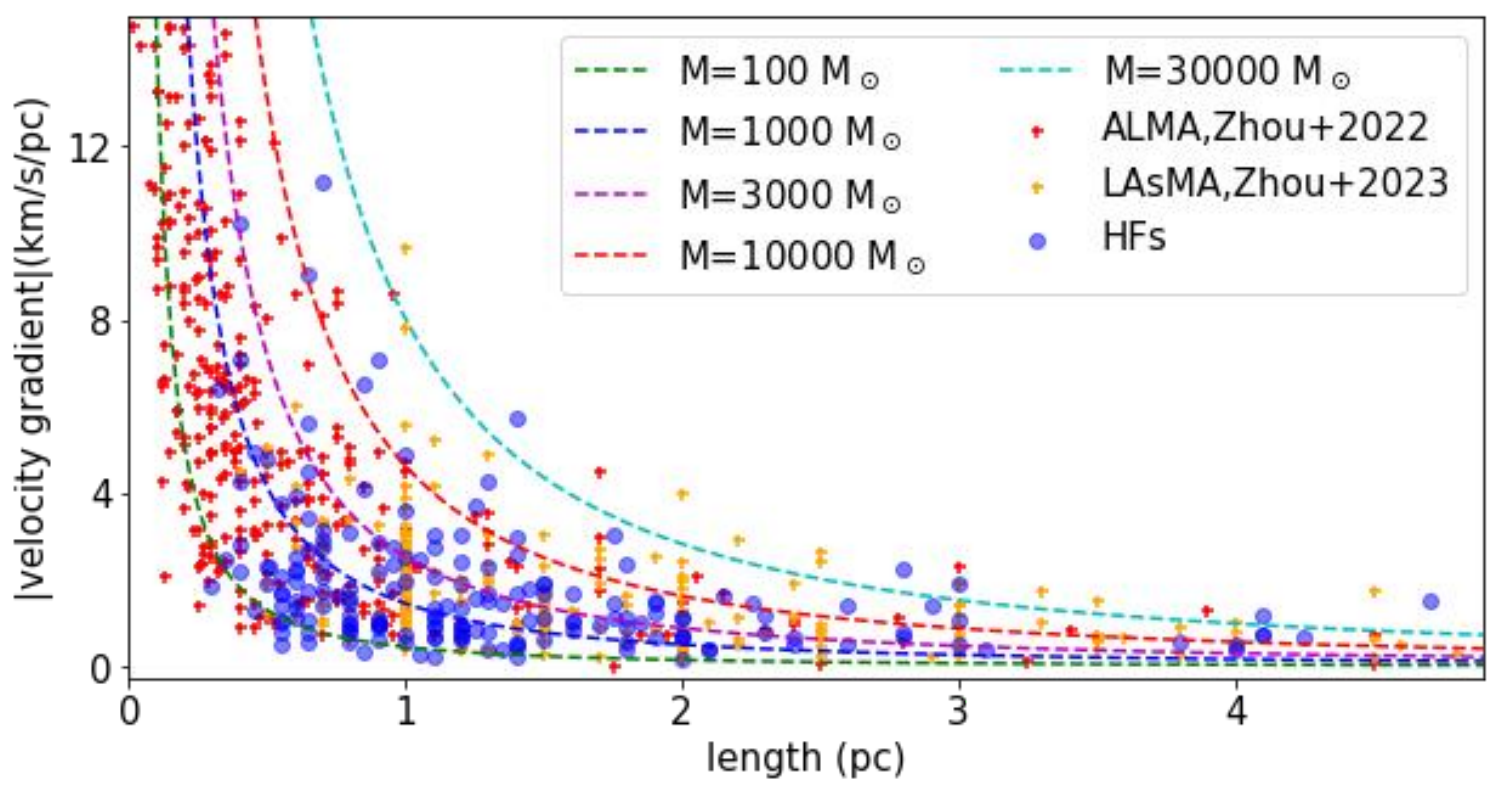}
\caption{Velocity gradient versus the length over which the gradient has been fitted. Red and orange "+" represent the velocity gradients fitted in \citet{Zhou2022-514} and \citet{Zhou2023-676}, respectively. The dashed lines show free-fall velocity gradients for comparison. For the free-fall model, from left to right, the lines denote masses of 100 M$_\odot$, 1000 M$_\odot$, 3000 M$_\odot$, 10000 M$_\odot$ and 30000 M$_\odot$, respectively.}
\label{gradient}
\end{figure}

A kinematic feature of hub-filament structures is that the gas flow along the filament converges towards the central hub, resulting in a measurable velocity gradient along the filament. Since we are now studying each local hub-filament structure, we follow the same analysis presentend in \citet{Zhou2022-514} to fit the velocity gradients around the intensity peaks. In Fig.~\ref{gradient}(a), the velocity gradients fitted in those local hub-filament structures are quite comparable to those presented in \citet{Zhou2022-514} and \citet{Zhou2023-676}. 
For the velocity gradients measured in this work, almost all of them can be fitted with masses in the range $\sim$ 100$-$10000 \,M$_\odot$, which is consistent with the mass distribution of leaf structures shown in Fig.\ref{compare}(b), indicating that local dense structures as gravitational centers can accrete the surrounding diffuse gas and produce the observed velocity gradients. 
Measurements of velocity gradients provide evidence for gas inflow within these structures, which can serve as kinematic evidence that these structures can be regarded as hub-filament systems.

\subsection{Physical
properties}\label{contrast}

\begin{figure*}
\centering
\includegraphics[width=1\textwidth]{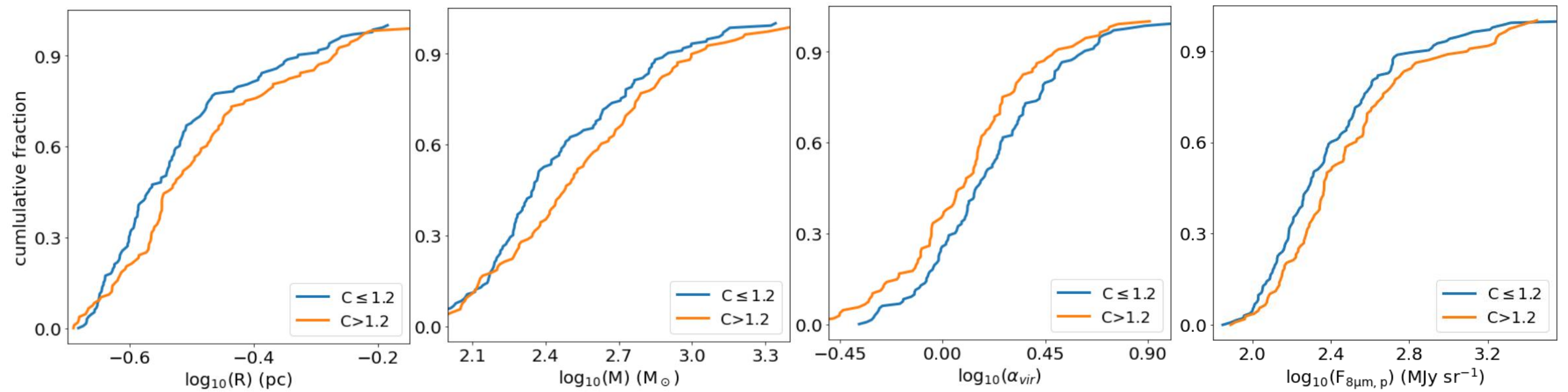}
\caption{Comparison of non-HFs with the density contrast $C$<1.2  and $C$>1.2. (a) Effective radius; (b) Mass; (c) Virial ratio; (d) Peak intensity of {\it Spitzer} 8 $\mu$m emission.}
\label{leaf-c}
\end{figure*}

\begin{figure}
\centering
\includegraphics[width=0.45\textwidth]{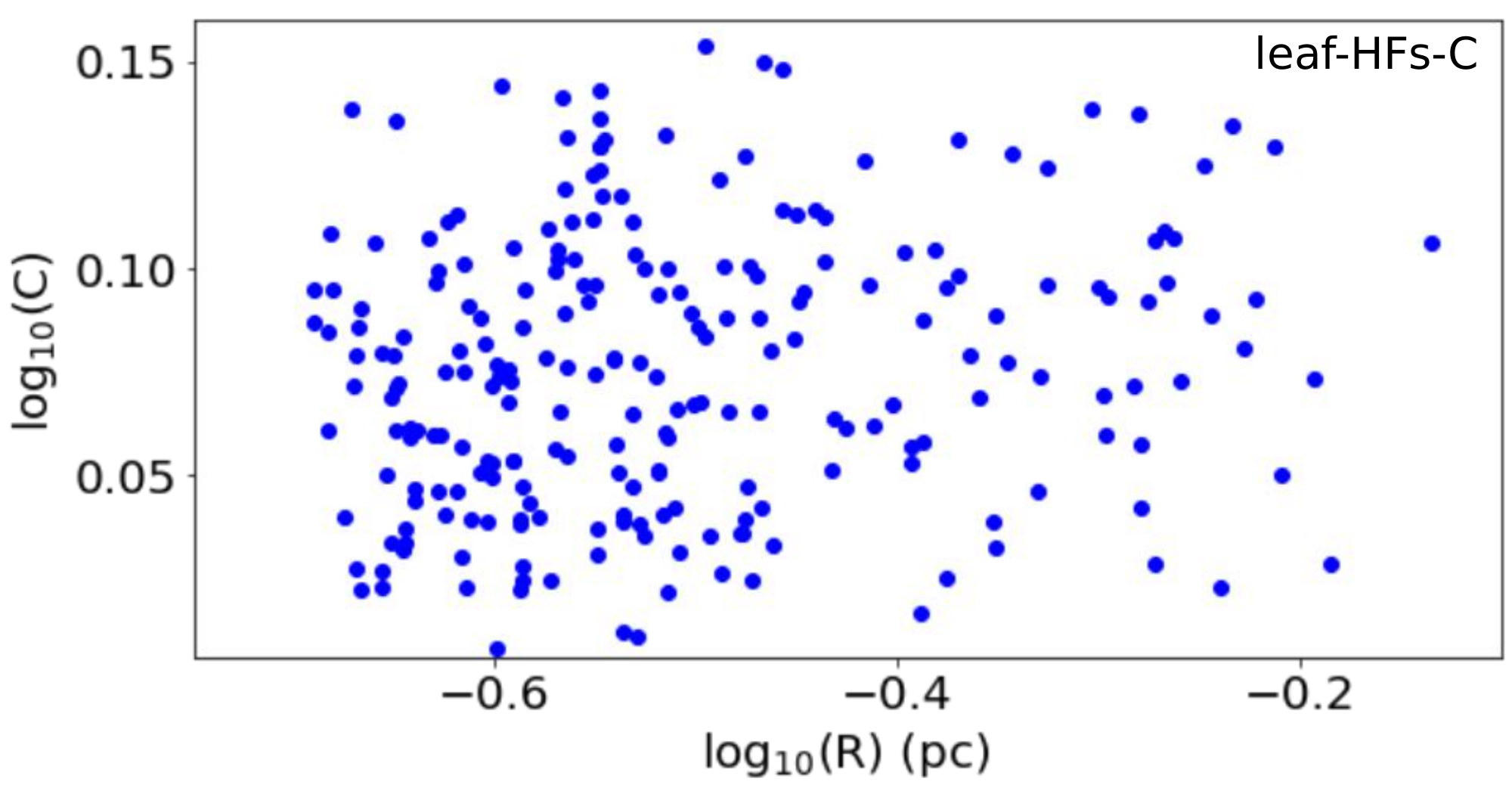}
\caption{The correlation between density contrast and scale of non-HFs.}
\label{smear}
\end{figure}

As expected, in Fig.~\ref{compare}(a), HFs display significantly higher density contrast. 
Moreover, HFs also possess the largest masses and the lowest virial ratios. Nearly half of the leaf structures are non-HFs. We divided non-HFs into two groups, using the median density contrast of non-HFs ($C$ = 1.2) as the threshold. In Fig.\ref{leaf-c}, the structures in the group with the higher density contrast have larger scale, larger mass and lower virial ratio.  
We note that the virial parameter $\alpha_{vir}$ is only a ratio of the gravitational energy ($E_{g}$) and the kinetic energy ($E_{k}$) in this work, which only provides relative measurements of gravitational strength.
The value of $\alpha_{vir}$ should not be a criterion for determining whether a structure undergoes gravitational collapse.  

In Fig.~\ref{compare}(a) and (d),
non-HFs have significantly smaller density contrasts and scales. Therefore,
one would argue that only when the structural scale is large enough, the density contrast can clearly manifest. In other words, the hub size of non-HFs could be too small to be resolved. Moreover, beam smearing effect may create an illusion of uniform density distribution. 
In order to check these possibilities, in Fig.~\ref{smear}, we fitted the correlation between density contrast and scale of non-HFs, and found that there is no clear correlation between them. Moreover, since HFs and non-HFs have significant scale overlap (see Fig.~\ref{compare}(d)), if HFs can discern the hubs, non-HFs should be able to as well. Therefore, relative to HFs, the absence of the hubs in non-HFs is real and not merely a resolution issue. 

\subsection{ATLASGAL clumps and radio sources}\label{assoc}

\begin{figure}
\centering
\includegraphics[width=0.35\textwidth]{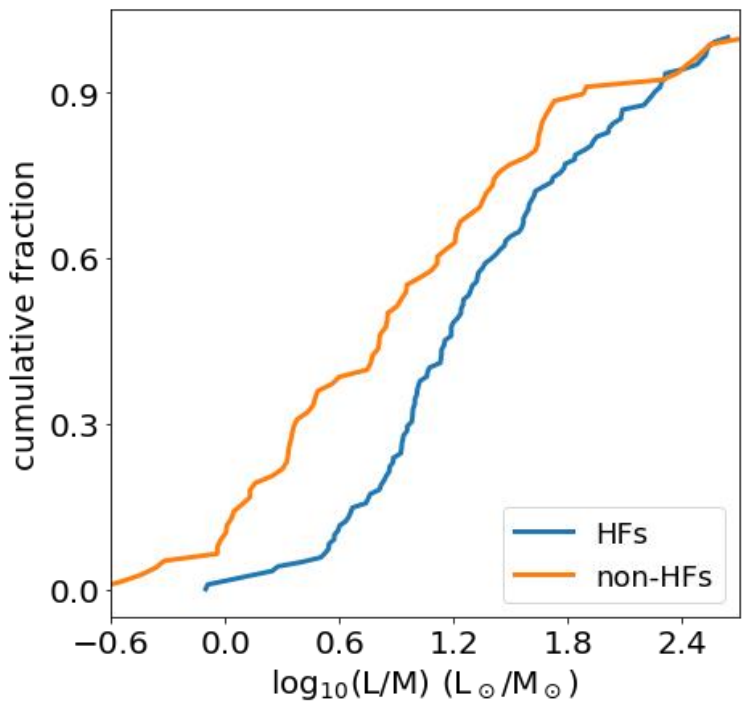}
\caption{The luminosity-to-mass ratio of ATLASGAL clumps associated with HFs and non-HFs.}
\label{sep}
\end{figure}

The physical properties of the ATLASGAL clumps \citep{Urquhart2022-510} associated with leaf structures can reveal their evolutionary stages. If the distance between the central coordinates of a clump and a leaf structure is less than the effective radius of the leaf structure, we associate the clump with the leaf structure.
In total, 438 leaf structures are
associated with 293 ATLASGAL clumps.
The fraction of hubs associated with ATLASGAL clumps is approximately 0.83 for HFs and about 0.32 for non-HFs. 
By the same way,
we also match leaf structures with the radio sources identified in the CORNISH-South survey \citep{Irabor2023-520}.
Approximately 29\% of hubs in HFs are associated with radio sources, compared to only around 4\% in non-HFs.
The ratio of HFs associated with clumps and radio sources is much higher than that of non-HFs, indicating that HFs are more evolved than non-HFs. In Fig.\ref{compare}(f),
the {\it Spitzer} 8 $\mu$m peak intensity of HFs is significantly higher than that of non-HFs.

The luminosity-to-mass ratios, $L/M$, of the clumps can be used to trace the evolutionary process of star formation \citep{Saraceno1996-309,Molinari2008-481,Liu2016-829,Molinari2016-826,Stephens2016-824,Elia2017-471,Urquhart2018-473,Elia2021-504,Traficante2023-520,Coletta2025-696}. 
In Fig.\ref{sep}, the clumps associated with HFs have significantly higher $L/M$ ratios compared to those associated with non-HFs, which also suggests that HFs are more evolved than non-HFs.

\section{Discussion}\label{discuss}

\subsection{Evolutionary sequence}

\begin{figure*}
\centering
\includegraphics[width=1\textwidth]{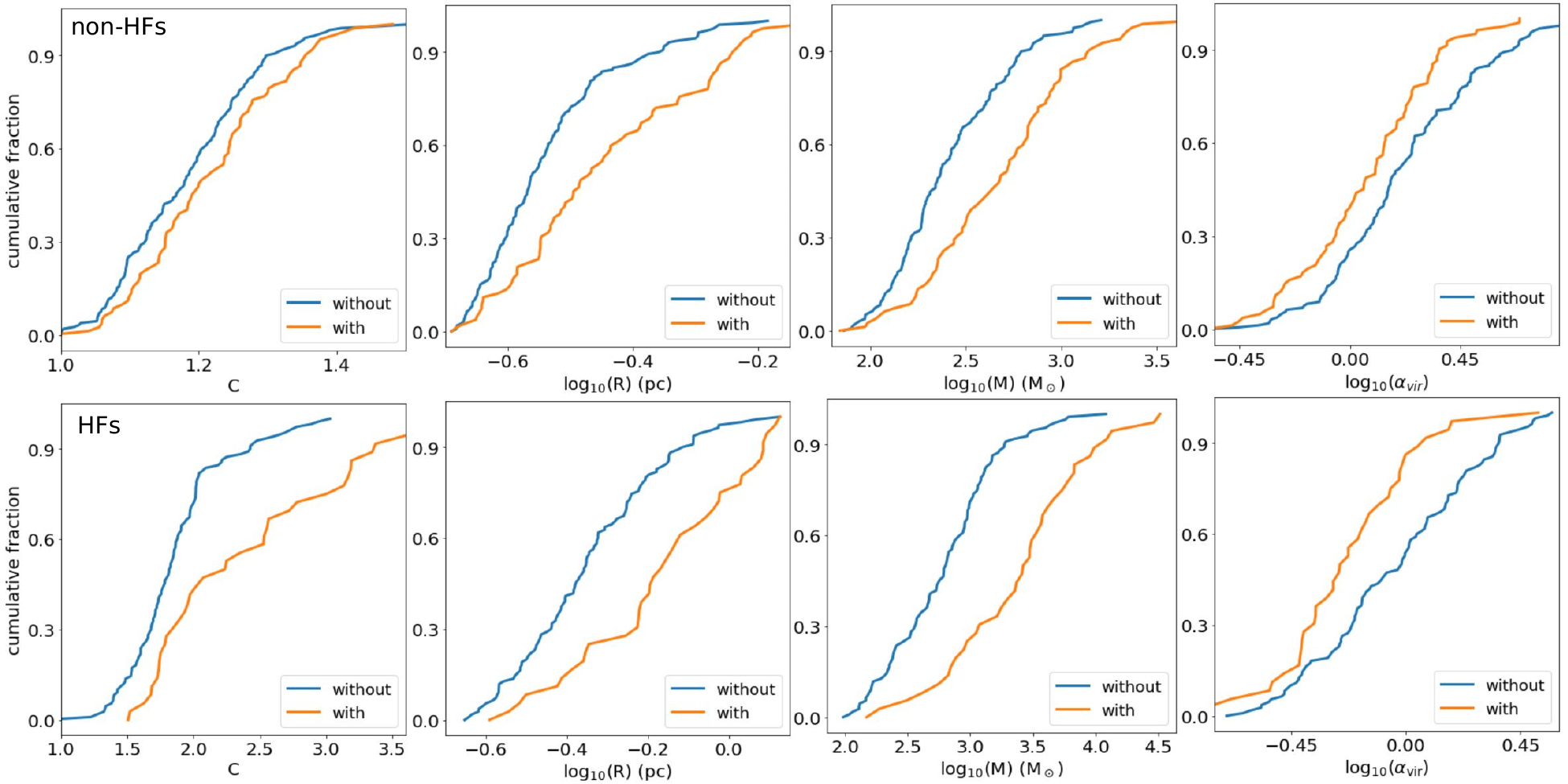}
\caption{The first row shows the physical properties of non-HFs without and with associated ATLASGAL clumps. For left to right, the physical parameters are the column density contrast defined in Sec.\ref{contrast}, effective radius, mass and virial ratio, respectively. The second row shows the physical properties of HFs without and with associated radio sources.}
\label{compare-AC}
\end{figure*}

In Fig.~\ref{compare-AC},
we compared the physical properties of two types of non-HFs, i.e. without and with associated ATLASGAL clumps. The structures associated with ATLASGAL clumps (more evolved) have higher density contrast, larger mass and lower virial ratio.
We also compared the physical properties of two types of HFs, i.e. without and with associated radio sources. The structures associated with radio sources (more evolved) have higher density contrast, larger mass and lower virial ratio. 
As confirmed in Sec.\ref{assoc}, HFs is more evolved than non-HFs. In Fig.\ref{compare}, compared with non-HFs, the more evolved HFs have higher density contrast, larger mass and lower virial ratio. Therefore, there may be an evolutionary sequence from non-HFs to HFs.
In Fig.\ref{large}, in terms of spatial distribution, HFs and non-HFs are mixed together, sharing the same physical environment. 
Therefore, the physical environments they inhabit may not be the main reason for the differences between HFs and non-HFs.
If there is an evolutionary relationship between them, these differences can be naturally explained.

\subsection{Density distribution}

As shown in Fig.\ref{contrast}(e),
although HFs have larger density contrast, the average density of non-HFs is almost comparable with HFs. Therefore,
non-HFs are not necessarily low-density structures. 
It is just that their density distribution is relatively uniform.
Leaf structures may undergo a gravitational focusing process starting from an initially relatively uniform density distribution, ultimately forming hub-filament structures. This process has been investigated in detail for the fragment G332.83-0.55 (one of HFs in this work) in \citet{Zhou2024-686}, where a strong gravitational center (hub) plays a crucial role in shaping the hub-filament morphology. According to the possible evolutionary sequence from non-HFs to HFs, currently, non-HFs lack a distinct gravitational focusing process that would result in significant density contrast. Thus, their density distribution is relatively uniform.
Therefore,
density contrast $C$ effectively measures the extent of gravitational collapse and the strength of the gravitational center of the structure, which definitively shape the hub-filament morphology. In terms of reflecting the development and evolutionary stage of a structure, density contrast is more crucial than the density itself.

\subsection{Structure of molecular clouds}

Considering the evolutionary sequence from non-HFs to HFs, non-HFs can be regarded as potential HFs, since they will evolve into HFs in the future. In this work, the molecular clouds are decomposed into hundreds of HFs and potential HFs with different scales. 
Combined with the kinematic evidence in \citet{Zhou2023-676} and \citet{Zhou2024-682-128}, we suggest that
molecular clouds are network structures formed by the gravitational coupling of multi-scale hub-filament structures. Hub-filament structures serve as the basic constituent units of molecular clouds. The knots in the networks are the hubs, and they are the local gravitational centers and the main star-forming sites. Actually, clumps in molecular clouds are equal to the hubs. 

This network picture can naturally explain the results in \citet{Zhou2024-682-128} and \citet{2024-682-173}.
In star-forming regions, spanning from cloud to core scales, initial-stage feedback has minimal impact on the physical properties of embedded dense gas structures. While feedback at these scales can disrupt the original cloud complex, the substructures within it can reorganize into new configurations governed by gravity around new gravitational centers. This dynamic process involves both structural destruction and formation, shifts in gravitational centers, but gravitational collapse is always ongoing.

The network structure of molecular clouds implies that the knots or local dense structures as local hubs or local gravitational centers are relatively independent of each other. 
Although some knots in the cloud evolve more rapidly, forming HII regions, due to their relative independence with the neighboring
knots, early feedback from them does not significantly impact the physical properties of the neighboring knots. Therefore, early feedback mainly reshapes the network structure of molecular clouds, representing a topological deformation rather than destruction of the molecular cloud, as illustrated in Fig.8 of \citet{2024-682-173}.

\section{Summary and conclusions}\label{summer}

We decomposed the G333 complex and the G331 GMC into multi-scale hub-filament structures using the high-resolution $^{13}$CO (3$-$2) data in LAsMA observation. The main conclusions are as follows:

1. The intensity peaks as the potential hubs were identified  based on the integrated intensity (Moment 0) map of $^{13}$CO (3$-$2) emission by the dendrogram algorithm. 
Around all intensity peaks, we extracted the average spectra in the 2.5 times extended regions to decompose their velocity components and fix the velocity range for each structure. The final identification of hub-filament systems (HFs) is based on the cleaned Moment-0 maps made by restricting to the fixed velocity ranges to exclude the possibility that the overlaps of uncorrelated velocity components produce fake hub-filament structures.

2. Based on morphology and the density contrast ($C$) value, we categorized all identified structures into two groups: those with a distinct high-density central region and those without. In general, a value of $C \approx 1.5$ serves as the boundary between these two categories. Ultimately, we identified 148 HFs and 243 non-HFs. Compared to non-HFs, HFs exhibit significantly higher density contrast, greater masses, and lower virial ratios.

3. We employed the filfinder algorithm to identify and characterize filaments within HFs based on integrated intensity maps. We extracted
the velocity and intensity profiles along the filaments, and fitted the velocity gradients around intensity peaks.
Changes in velocity gradient with scale roughly follow the free-fall model in the mass range $\sim$ 100--10000 \,M$_\odot$, which is consistent with the mass distribution of leaf structures, indicating that local dense structures as gravitational centers accrete the surrounding diffuse gas and produce the observed velocity gradients. Velocity gradient measurements provide evidence of gas inflow within these structures, offering kinematic support that they are hub-filament systems.

4. The ratio of HFs associated with clumps and radio sources is much higher than that of non-HFs. The {\it Spitzer} 8 $\mu$m peak intensity of HFs is significantly higher than that of non-HFs. The luminosity-to-mass ratios, $L/M$, of the associated ATLASGAL clumps  from HFs to non-HFs is decreasing. These results suggest that HFs are more evolved than non-HFs.

5. In non-HFs, structures associated with clumps exhibit higher density contrast, larger mass, and lower virial ratios compared to those without clump associations. Similarly, in HFs, structures linked to radio sources show higher density contrast, larger mass, and lower virial ratios than those without such associations. Additionally, compared to non-HFs, the more evolved HFs demonstrate higher density contrast, larger mass, and lower virial ratios. These observations suggest a potential evolutionary sequence from non-HFs to HFs.

6. Although HFs have larger density contrast, the average density of non-HFs is almost comparable with HFs. Thus, non-HFs are not necessarily low-density structures.  It is just that their density distribution is relatively uniform. According to the possible evolutionary sequence from non-HFs to HFs, currently, non-HFs lack a distinct gravitational focusing process that would result in significant density contrast. The density contrast $C$ effectively measures the extent of gravitational collapse and the strength of the gravitational center of the structure that definitively shape the hub-filament morphology. In terms of reflecting the development and evolutionary stage of a structure, density contrast is more crucial than density itself.

7. Molecular clouds are decomposed into hundreds of hub-filament structures of varying scales. 
We propose that molecular clouds are network-like structures formed through the gravitational coupling of multi-scale hub-filament structures. These hub-filament structures act as the fundamental building blocks of molecular clouds. The hubs serve as the nodes of the network, acting as local gravitational centers and primary sites of star formation.
The network structure of molecular clouds provides a natural explanation for why feedback from protoclusters does not significantly alter the kinematic properties of the surrounding embedded dense gas structures, as concluded in our previous studies.

\begin{acknowledgements}
We would like to thank the referee for the detailed comments and suggestions, which have helped to improve and clarify this work.
Thanks to E. Vázquez-Semadeni and S. H. Li for providing helpful comments.
This publication is based on data acquired with the Atacama Pathfinder Experiment (APEX) under programme ID M-0109.F-9514A-2022. APEX has been a collaboration between the Max-Planck-Institut für Radioastronomie, the European Southern Observatory, and the Onsala Space Observatory.
Tie Liu acknowledges the supports by the National Key R\&D Program of China (No. 2022YFA1603101), National Natural Science Foundation of China (NSFC) through grants No.12073061 and No.12122307, the PIFI program of Chinese Academy of Sciences through grant No. 2025PG0009, and the Tianchi Talent Program of Xinjiang
Uygur Autonomous Region.
MJ acknowledges the support of the Research Council of Finland Grant No. 348342. 
This research made use of astrodendro, a Python package to compute dendrograms of Astronomical data (http://www.dendrograms.org/).  
\end{acknowledgements}


\bibliography{ref}
\bibliographystyle{aasjournal}

\begin{appendix}
\twocolumn

\end{appendix}

\end{document}